\documentstyle[sprocl]{article}

\def\PLB{{\em Phys. Lett.} B}
\def\PRL{\em Phys. Rev. Lett.}
\def\PRD{{\em Phys. Rev.} D}

\def\OMC{$\cal O \ $}
\def\OMCN{$\cal O$}

\def\ra{\rightarrow}
\def\be{\begin{equation}}
\def\ee{\end{equation}}
\def\bea{\begin{eqnarray}}
\def\eea{\end{eqnarray}}
\begin{document}

\title{ASPECTS OF CHARMONIUM}

\author{  S.F. TUAN}
\address{Department of Physics, University of Hawaii at Manoa \\
2505 Correa Road, Honolulu, HI 96822, U.S.A.}

\maketitle\abstracts{Here we discuss three separate topics of charmonium.
           (1) The Omicron \OMC as $1^{--}$ trigluonia, and the charmonium
           puzzle; the $\eta_c' (3600)$ search. (2) The decay $\psi(^1P_1)
           \ra \pi^o + J/\psi$ and isospin violating charmonium decays.
           (3) How to search for $\chi_c' = \chi_c(2P)$ charmonium states.}
\noindent
{\bf Introduction}

\noindent
I wish to thank the Chair of the Organizing Committee of Hadron '95,
Sandy Donnachie, for choosing for me the topic of my talk on {\bf Aspects
of Charmonium}. Such an umbrella choice enables me to cover three disparate
topics in which I have done work in association with others over the past
couple of years. The common denominator being that at least the word
``charmonium" (as well as, by and large the content), figures prominently
in all three topics. Sections 1, 2, and 3 below study these topics with
conclusions/summaries attached to each separately. So here we go!

\section{\bf Omicron \OMC as $1^{--}$ Trigluonia, and the Charmonium Puzzle.}

First the name Omicron, for short \OMCN, is given to a gluonium/glueball state
with $(I,J^{PC}) = (0,1^{--})$ made up from three gluons $3g$ (hence
trigluonia).
The \OMC can of course decay into exclusive light hadron channels h, according
to the scheme $3g \ra$ \OMC $\ra h$. In earlier times \cite{hou}
theoretical estimates for
the mass of \OMC hover around 2.4-2.5 GeV, though some other estimates$^2$
suggested that it could be as high as 3 GeV. Hou and Soni \cite{hou}
argued further that
the interpretation of a $1^{--}$ \OMC is easy and has
definite signature advantages
over glueballs with $J^{PC} = 0^{-+}, 0^{++}$, and $2^{++}$. For
instance (a) the mixing
of  \OMC with ordinary quarkonia is small 0$(\alpha_s^3)$ to
$0(\alpha_s^2)$, (b) \OMC $\ra e^+e^-$ decay width $\sim 0.3$ eV, and
$BR($ \OMC $\ra e^+e^-) \ \sim 10^{-8}$,
while (c) $e^+e^-$
production of Omicron via intermediate $\gamma^*$ has a cross section
$< 20pb$ - exceedingly small when compared with the corresponding
case of quarkonium
production. The search method for \OMC in charmonium could deploy the following
decay processes:-

\begin{eqnarray*}
J/\psi \ra \pi^+\pi^- {\cal{O}} (\ra \rho \pi), \ \mbox{for} \ {\cal{O}} \
\mbox{mass} < 2.7 \ GeV   
     \\
\psi' \ra \pi^+\pi^- {\cal{O}} (\ra \rho \pi),  \ \mbox{for} \ {\cal{O}} \
\mbox{mass} < \ 3.3 \ GeV.
\end{eqnarray*}
Though the \OMC exists in its own right, attempts have been made to relate
it to the outstanding puzzle of charmonium physics, namely the absence
of $\psi'$
decays to vector-pseudoscalar $VP$ light hadrons. On the basis of perturbative
QCD, it is reasonable to expect that for any hadronic final state $h$, we have
\be
{\cal{Q}}_{h} \equiv \frac{B (\psi^\prime \ra h)}{B(J/\psi \ra h)}
       \cong  \frac{B (\psi^\prime \ra e^+e^-)}
                   {B (\psi^\prime \ra e^+e^-)}
       = (14 \pm 2) \%
\ee
\noindent
This is well documented \cite{rev} for
$p\bar{p}\pi^o, 2\pi^+2 \pi^-\pi^o,
2 (\pi^+ \pi^-), \pi^+\pi^-
p\bar{p}$,
$3\pi^+3 \pi^- \pi^o,
\linebreak
b_1^\mp (1230)\pi^\mp$
(from recent BES data) hadronic
channels. The startling exceptions occur for $VP=\rho\pi,
K^*\bar{K} +c.c$. where the
present experimental limits are $Q_{\rho\pi} < 0.0048$,
$Q_{K^{*}\bar{K} + c.c}
< 0.0108$
(where the factor of 3 correction to PDG (1994) \cite{rev} due
to Y.F. Gu, acknowledged
by M. Roos, is taken into account). Brodsky-Lepage-Tuan \cite{brod},
henceforth referred
to as BLT, proposed a coherent explanation of the puzzle by assuming that the
violation of the perturbative QCD theorem \cite{bro} that total hadron
helicity is conserved in high-momentum-transfer exclusive processes
when $J/\psi$
decays to hadrons via three hard gluons is modulated by
the gluons forming an intermediate
gluonium $1^{--}$ state, \OMCN, before transition to hadrons $h$.
It is the \OMC (which has
large transverse size) which does not respect the helicity theorem,
and accounts for the relatively large branching ratios $(\sim 10^{-2})$ for
$J/\psi \ra \rho \pi, K^*\bar{K}+c.c.$
To account for the constraints
$Q_{VP}$, the \OMC has a mass within 80
MeV of the $J/\psi$ mass, and total width $< 250$ MeV (using latest
$VP=K^*\bar{K}+c.c.$
constraint). We need to note a more recent work by Anselmino
et al. \cite{ans} who have
suggested that the BLT assumptions can be met with a more modest Omicron
 width,
to wit $\Gamma_{tot}$(\OMCN) between 10 to 100 MeV.

The problems with the BLT/Anselmino et al. solution are (i) an \OMC at 3 GeV
(unlike the $\omega^o)$ has many decay channels, so why should
\OMC $\ra$ VP dominate amongst decays of
MeV of the $J/\psi$ mass, and total width $< 250$ MeV (using latest
$VP=K^*\bar{K}+c.c.$
constraint). We need to note a more recent work by Anselmino
et al. \cite{ans} who have
suggested that the BLT assumptions can be met with a more
modest Omicron width,
to wit $\Gamma_{tot}$(\OMCN) between 10 to 100 MeV.
     The problems with the BLT/Anselmino et al. solution are (i) an \OMC at 3
GeV
(unlike the $\omega^o)$ has many decay channels, so why should \OMCN $\ra$ VP
dominate amongst decays of \OMCN. (ii) The mystery of the electromagnetic
$J/\psi \ra \omega \pi^o \ VP$ decay. Here $\Gamma(J/\psi
\ra \omega \pi^o) > 3 \Gamma (J/\psi \ra \pi^+ \pi^-)$ with a
$BR(J/\psi \ra \omega \pi^o) = (4.2 \pm 0.6) \times 10^{-4}$. As a pure
$I=1$ transition via virtual
$\gamma^*$ with $q^2>0$, the Omicron cannot be involved,
yet the $VP$ mode here is
three times larger than the non $VP$ mode $h=\pi^+\pi^-!$
An intriguing experimental
measurement, if hadron helicity conservation HHC theorem~\cite{bro}
is {\bf not} applicable
at the $J/\psi-\psi'$ mass range, is to measure $\psi' \ra
\omega \pi^o$ at the level given
by the 14\% rule of Eq. (1), i.e. 0.14 $BR(J/\psi \ra \omega
\pi^o) \sim 0.6 \times
10^{-4}$. If found,
one can infer that the Brodsky-Lepage HHC theorem \cite{bro}
is {\bf not} valid even at the
$\psi'$ mass scale! (iii) The BES Collaboration studied the
vector-tensor $VT$ case
$\psi' \ra \omega f_2^0 (1270)$ recently.
Now this decay is {\bf allowed} by HHC theorem \cite{gu}, yet
the experimental finding is that it is suppressed with
$Q_{\omega f_{2}^0} < $2.6\%. Hence
the QCD HHC Theorem~\cite{bro} appears to be not relevant at the $\psi'$ mass
scale for $h=VT$
also, when evaluated in terms of the charmonium puzzle.
(iv) The $\rho-\pi$
channel has been carefully scanned across the $J/\psi$ region in
$e^+e^-$ annihilation at BES \cite{che} and there has been no sign
of distortion of $J/\psi \ra \rho\pi$ shape
due to Breit-Wigner \OMC interference. I am myself particularly impressed by
the
upper limit on the total \OMC width of $< 8 $MeV set by the
BES Collaboration~\cite{ols}.
(v) Finally there is the work of the UKQCD Collaboration~\cite{bal}
on SU(3) Glueballs-Comprehensive Lattice Study. This predicts
the \OMC $(1^{--})$ at
4 GeV and optimists
in the field believe the prediction could be good up to 100 MeV. At NATO's
Advanced Study Institute at Swansea, Chris Michael assured me that he is very
sure that the \OMC $(1^{--})$ gluonium is $> 3$ GeV.

     A certain number of conclusions can now be drawn. First, the HHC
Theorem \cite{bro}
is probably {\bf not} relevant at the mass scale
of $J/\psi[\eta_c(2980)]$ nor at
$\psi'[\eta_c'(3600)]$. Secondly, the BLT model \cite{brod}
including recent update \cite{ans} for
solution of $J/\psi(\psi')\ra \rho\pi,K^*\bar{K}$ puzzle and
Anselmino et al.\cite{an} model
predicting an $0^{-+}$ trigluonia near $\eta_c(2980)$ are probably wrong,
since they
both assume the validity of the HHC Theorem of Brodsky-Lepage \cite{bro}.
Thirdly, the
puzzle of the suppression of $\psi' \ra VP,VT$ channels remains an
open problem in need of a solution.

     I wish however to point out that the study of the charmonium puzzle has
led to the discovery of the following useful relationship \cite{cha}
\be
          BR(\eta_c' \ra h) \sim \ BR(\eta_c  \ra h)
\ee
where $h$ is an exclusive final state. Since the $\eta_c(2980)$
has known hefty
decays into $K\bar{K}\pi, \eta\pi \pi, \eta' \pi\pi$, the use of
(2) could enable one to
search for $\eta_c'(3600)$ directly via the chain $\psi' \ra
\gamma \eta_c' \ra \gamma h$,
where $\eta_c'\ra h$ of the last leg is estimated using (2). For a
$5 \times 10^6 \ \psi'$
sample, some 20,000 $\eta_c'$ events can be expected, and the breakdown into
various exclusive modes $h$ are given elsewhere \cite{tua}.
If IHEP-Beijing were to
acquire a crystal-barrel type detector with high efficiency for detecting
low energy $\gamma$ from the $\psi' \ra \gamma \eta_c'$ leg,
this could be a feasible
search method. For the $\bar{\mbox{p}}\mbox{p}$ process,
the peak cross section for
$\bar{\mbox{p}}\mbox{p}\ra
\eta_c'(3600)\ra h$ using (2) can be expected to be substantially
larger \cite{tua}
than that even for $\bar{\mbox{p}} \mbox{p}\ra
\eta_c(2980) \ra 2 \gamma$. If FERMILAB
were to acquire
a magnetic spectrometer detector allowing for clean detection
of $K\bar{K} \pi$
and other exclusive modes $h$, this could be a feasible search method for
$\eta_c'$. Finally Sergei Sadovsky has suggested at this meeting that the
$\eta_c'$  be searched for in two photon physics $e^+e^- \ra e^+e^-
\gamma \gamma \ra \eta_c'$, where again the relation (2) can be kept
in mind in estimating event rates.

\section{\bf The $\psi(^1P_1) \ra \pi^o J/\psi$ and Isospin Violating
Charmonium Decays.}

     The recent observation of the isospin-forbidden decay of the $^1P_1$
charmonium state revives interest in these pionic decays of
 charmonium \cite{tuan}.
For instance E760 found

\be
   \frac{BR(\psi(^1P_1) \ra J/\psi+\pi^o)}
         {BR(\psi(^1P_1) \ra J/\psi+\pi\pi)}
   > 5.5 \ (90\% \ C.L.).
\ee
Why is the isospin violating $\psi(^1P_1) \ra J/\psi+\pi^o$
decay so large compared
with the strong isospin conserving decay $\psi(^1P_1) \ra J/\psi + \pi\pi?$
This certainly violates theoretical and intuitive understanding
(Misha Voloshin excepted). For instance back in 1979,
Isgur et al. \cite{isg} estimated that
$BR(\psi(^1P_1) \ra J/\psi+\pi^o) \sim 10^{-3}$ to $10^{-4}$,
while the canonical estimate \cite{kua}
for the $BR(\psi(^1P_1) \ra J/\psi+\pi\pi)$ is of order $10^{-2}$.
The simplest explanation
of the anomaly (3) is that since E760 also did {\bf not} see $\psi(^1P_1)
\ra \gamma \eta_c$
which is very much the dominant mode with branching ratio $\sim 50\%$ by
any reasonable estimate, we really need to await E835
next year to confirm this
``discovery" of $\psi(^1P_1)$ at 3526 MeV - via this peculiar isospin
violating mode involving single pion emission, while the isospin
allowed dipion mode and the
dominant $\gamma \eta_c$ mode remain to be identified.
     Nevertheless Lipkin and I noted \cite{lip} that in fact
SU(3)/isospin violating
charmonium decays do show a pattern of being anomalously large, e.g.
$BR(\psi' \ra J/\psi \eta^o) = 2.7 \pm 0.4\%$ [SU(3) violating] and
$BR(\psi' \ra J/\psi \pi^o) = (9.7 \pm 2.1) \times 10^{-4}$
[isospin violating]. Maybe $\psi(^1P_1) \ra J/\psi+\pi^o$
is just
following this tradition? First from the above data, and
taking a p \cite{rev} phase
space dependence appropriate to the P-wave decays involved, we have
\be
          BR(\psi' \ra J/\psi+\eta)/ \mbox{phase space} \  = 3.375  
\label{eq:sira}
\ee
\be
BR(\psi' \ra J/\psi+\pi^o)/ \mbox{phase space} = 6.6 \times
      10^{-3}.       
\label{eq:sisi}
\ee

Both ($\ref{eq:sira}$) and ($\ref{eq:sisi}$) violate
SU(3) and OZI-rule. However ($\ref{eq:sisi}$) in addition violates
isospin as well. Hence in terms of magnitude comparison, if we regard isospin
violation as electromagnetic, we need to compare $[(\alpha/\pi)$ or
$\alpha]^2 \times$
Eq. (\ref{eq:sira}) with Eq.(\ref{eq:sisi}), i.e. $1.69 \times 10^{-5}$
to $1.69\times 10^{-4}$ with $6.6 \times 10^{-3}$. Hence the
pionic emission appears to dominate $\eta^o$ emission by a
factor of 400 to 40!
(Parenthetically the $\psi' \ra J/\psi+\eta^o$ decay may also
be somewhat enhanced
relative to SU(3) allowed $\psi' \ra J/\psi+\pi+\pi.)$
The strength of (5) thus tends
to support a non-electromagnetic contribution to this decay as well.

We construct a model by relating $\psi' \ra J/\psi+\pi^o$ to $\psi'
\ra J/\psi+\eta^o$
by using constituent-quark phenomenology rather than hadron field theory. The
decay is described by a ``black box" model in which the charmonium
$\psi'$ initial state enters and from which the final charmonium state
$J/\psi$ emerges together with a quark-antiquark pair which hadronizes
into a physical pseudoscalar meson $(\pi^o, \eta^o$). All processes in the
 black box are described by
phenomenological parameters determined by fitting experimental data.
The relevant suppression factor for comparing these two
forbidden processes (4) and
(5) is not determined by $\alpha$ but rather by the ratio of the two symmetry
breaking parameters that mix an SU(3) into the $\pi^o$ and $\eta^o$
wave functions.
This is just the ratio of the two mass differences that characterize
the relevant symmetry breaking; e.g $(m_d - m_u)/(m_s - m_u)$. The quantity
is thus considerably larger than $\alpha$. There is also additional symmetry
breaking which
is purely electromagnetic. We have included this in our model
where it is characterized by the ratio
$\{M (\pi^\pm)-M(\pi^o)\}/\{M(\eta)-M(\pi^o)\}$ which is closer
to $\alpha$ but still larger than $\alpha$ by a factor $\sim 1.5$. These two
terms add
coherently and enhance the isospin violating
$\psi' \ra \pi^o J/\psi$ even more. In the actual analysis
the term proportional to the quark mass difference is a bit more
than double the electromagnetic term so that the square of the sum is an order
of magnitude larger than the square of the electromagnetic term and therefore
between one and two orders of magnitude larger than $\alpha$ \cite{cas} -
as needed for
understanding (4) and (5). To summarize, our black box model for flavor
symmetry breaking gives reasonably good results in agreement with experiment if
we take $m_d - m_u = 3.33$ MeV, a value which seems to be in the right ball
park.  For $\psi (^1P_1)$ at 3526 MeV$^{14}$ only the pionic decay
is available, because $\psi(^1P_1) \ra \ J/\psi + \eta$ is phase
space forbidden.  Nevertheless from orbital angular momentum consideration plus
the spirit of our discussion above,  we expect isospin forbidden
$\psi(^1P_1) \ra J/\psi + \pi^0$ could be much enhanced compared with allowed
$\psi(^1P_1) \ra J/\psi + \pi \pi$.

\section{\bf How to Search for $\chi_c' = \chi_c(2P)$ Charmonium States.}

     We shall concentrate on the $J=1, J=2$ members since the $J=0$ member
is expected to be broad. The problem with the radially excited
$\chi_c$-states is that
theoretical estimates \cite{god} say that their masses are 3950 MeV
(for $J=1)$ and 3980
MeV (for $J=2)$ and hence lie above the $D\bar{D}$ threshold.
However it has been argued that at least for $J=2$,
the $\chi_{c2}(2P) \ra  D\bar{D}, D\bar{D}^*$
proceed via orbital $L=2$ and there could be significant
suppression of these D-wave decays due
to limited phase space provided that the mass estimate is not significantly
larger than that given by theory. Hence it is possible that $\chi_{c2}(2P)$ has
significant branching ratio to charmonium states below the $D\bar{D}$
threshold.
Philip Page \cite{pag} has calculated the total widths of the $\chi_c'$ states,
 and found
that it is very probable that the $J=2$ member has a total width between 1 to
5 MeV. So the dominant decay could well be $\chi_{c2}' \ra \gamma+\psi'$
below $D\bar{D}$
threshold. Back of the envelope type calculation would suggest that
the branching ratio for this decay could well be in
the range 1-10\%, and for illustration we shall take
$BR(\chi'_{c2} \ra \gamma+\psi') = 1, 5, $10\%.

The search method for $\chi_{cJ}'$ states has been described in some detail
elsewhere \cite{tu}, so we shall here just summarize the
results for the $J=2$ case.
At CLEO II, we take advantage of inclusive decays of B-mesons to charmonium.
In particular we estimate $B_J = BR(\chi'_{cJ} \ra \gamma+ \psi') \times
BR(B \ra \chi_{cJ}'+X)$. The
CLEO II measurement for $BR(B\ra\chi_{c2}+X)$ is $0.25 \pm 0.10\%$.
The unknown $BR(B \ra
\chi'_{c2}+X)$ is then estimated by introducing the multiplicative
Braaten correction factor \cite{tu} of 1.32. Hence taking the central
value for $BR(B\ra\chi_{c2}+X)$
we find $B_2 = 3.30 \times 10^{-5}, 1.65 \times 10^{-4}, 3.30 \times
10^{-4}$ for
$BR(\chi'_{c2} \ra \gamma + \psi')$
= 1,5,10\% respectively. To estimate the number of
observed J=2 events, $N_J^{obs}$,
we need the number of B-mesons produced on a good year from
$e^+e^- \ra \Upsilon (4S)
\ra B\bar{B}$ (where most of the $B\bar{B}$ events are collected)
multiplied by $B_J$ and $BR(\psi' \ra J/\psi \pi^+\pi^-) \times
{\Sigma_{l= \mu,e}} \
BR(J/\psi \ra l^+l^-) \times \epsilon$, where
$\epsilon$ is the efficiency for detecting $\psi' \ra J/\psi \pi^+\pi^-$
in the $J/\psi$
dilepton mode (about 20\% at CLEO II).  Collecting the pieces
together, we have
 \be
         N_2^{obs} = 1.09, 5.43, 10.86  \ \mbox{events}
\ee
for $BR(\chi'_{c2} \ra \gamma+\psi') = 1,5,10\%$
respectively.

     Our conclusion is that optimistically we can expect $N_2^{obs}$
in the ball
park of 10 events per year. They can be steadily increased by extending
$B\bar{B}$
accumulation over a period of {\bf several} years.
We remark however that the CDF
$\bar{p} p$ process has already identified $\chi_{cJ}$ from $\gamma-J/\psi$
mass spectrum.
Hence the invariant mass spectrum of $\gamma-\psi'$
can be studied also at the
predicted mass values for $\chi'_{cJ}$.
Given the large statistics possible at CDF,
they could well beat out CLEO II in the $\chi'_{cJ}$ search!

\section*{\bf Acknowledgments}

     I would like to thank Professor David V. Bugg for a scholarship at the
NATO Advanced Study Institute in London and Swansea where portions of this
work was completed. This work was supported in part by the US Department of
Energy under Grant DE-FG-03-94ER40833.

\end{document}